\documentclass[a4paper,twoside]{article}

\usepackage{epsfig}
\usepackage{subcaption}
\usepackage{calc}
\usepackage{amssymb}
\usepackage{amstext}
\usepackage{amsmath}
\usepackage{amsthm}
\usepackage{multicol}
\usepackage{pslatex}

\usepackage{natbib} 
\let\bibhang\relax
\usepackage{apalike}

\usepackage{csquotes}
\usepackage{xcolor}
\usepackage[export]{adjustbox}

\usepackage{SCITEPRESS}     

\begin{document}

\title{Perception and Acceptance of an Autonomous Refactoring Bot}

\author{
\authorname{Marvin Wyrich\sup{1}\orcidAuthor{0000-0001-8506-3294},
Regina Hebig\sup{2}\orcidAuthor{0000-0002-1459-2081},
Stefan Wagner\sup{1}\orcidAuthor{0000-0002-5256-8429}
and Riccardo Scandariato\sup{2}\orcidAuthor{0000-0003-3591-7671}}
\affiliation{\sup{1}University of Stuttgart, Germany}
\affiliation{\sup{2}Chalmers $\vert$ University of Gothenburg, Sweden}
\email{\{marvin.wyrich, stefan.wagner\}@iste.uni-stuttgart.de, \{regina.hebig, riccardo.scandariato\}@cse.gu.se}
}

\keywords{Software Bot, Refactoring, Human Agent Interaction, Collaborative Development, Software Engineering}

\abstract{The use of autonomous bots for automatic support in software development tasks is increasing.
In the past, however, they were not always perceived positively and sometimes experienced a negative bias compared to their human counterparts.
We conducted a qualitative study in which we deployed an autonomous refactoring bot for 41 days in a student software development project.
In between and at the end, we conducted semi-structured interviews to find out how developers perceive the bot and whether they are more or less critical when reviewing the contributions of a bot compared to human contributions.
Our findings show that the bot was perceived as a useful and unobtrusive contributor, and developers were no more critical of it than they were about their human colleagues, but only a few team members felt responsible for the bot.
}

\onecolumn \maketitle \normalsize \setcounter{footnote}{0} \vfill

\newcommand{\researchQuestionOne}{RQ1: How do developers perceive the participation of a refactoring bot in their project?}
\newcommand{\researchQuestionTwo}{RQ2: Are developers more or less critical when reviewing the contributions of a refactoring bot compared to human contributions?}
\newcommand{\researchQuestionThree}{RQ3: How do developers think an autonomous refactoring bot should ideally be designed?}

\section{\uppercase{Introduction}}
\label{sec:introduction}

Refactoring has been defined as \enquote{the process of changing a software system in such a way that it does not alter the external behavior of the code yet improves its internal structure}~\cite[p. xvi]{Fowler1999Refactoring:Code}.
It is essential to continuously go through this process to improve the quality and maintainability of the source code, thereby increasing the productivity of developers~\citep{Moser2008ATeam} and avoiding the accumulation of technical debt in the system~\citep{Avgeriou2016}.
Some static code analysis tools identify refactoring opportunities in the form of \textit{code smells}, which are functioning program code that is poorly structured~\citep{Fowler1999Refactoring:Code}.
Manually removing these code smells is error-prone, tedious and sometimes challenging~\citep{Bavota2012WhenStudy,Kim2011AnEvolution,Kim2012ABenefits}.
The effort and associated costs may also become too high to be justified to a client or project manager, which prevents developers from having the necessary resources or organizational support to manually improve the quality of the source code~\citep{Yamashita2013DoSurvey}.

To make the removal of code smells more efficient and more effective, \cite{Wyrich:2019:RefactoringBot} implemented an autonomous bot that automatically refactors code and submits its changes to the development team for asynchronous review in the form of pull requests.
A recent study has shown that such changes with automatically fixed code smells are generally accepted by developers~\citep{marcilio2019automatically}.
However, pull requests in that study were proposed manually and only a single time after prior consultation with the project maintainers.
Furthermore, we know that contributions are not only evaluated on their content, but also on the social characteristics of the contributor~\citep{terrell2017gender,Ford:2019:BeyondCode}.
In the case of contributing bots, identifying them as bots can be sufficient to observe a negative bias compared to contributions from humans~\citep{Murgia2016AmongWebsites}.

We therefore introduced the \textit{Refactoring-Bot} in a software development team and had it continuously contribute refactoring suggestions for 41 days.
The developers knew it was a bot and could interact with it.
In between and at the end, we conducted semi-structured interviews to answer the following research questions:

\begin{itemize}
    \item \researchQuestionOne
    \item \researchQuestionTwo
    \item \researchQuestionThree
\end{itemize}

\section{\uppercase{Related Work}}
\label{sec:relatedwork}

\cite{Wyrich:2019:RefactoringBot} describe the \textit{Refactoring-Bot} as \enquote{an autonomous bot that integrates into the team like a human developer via the existing version control platform}.
It currently supports a handful of refactoring operations to eliminate code smells reported by the static code analysis tool SonarQube.
These operations are removing unused method parameters, unused private fields and commented-out code, correcting the wrong order of modifiers, adding missing override annotations and immediately returning an expression instead of assigning it to a new variable.
It is also possible to interact with the bot via comments in its pull requests, for example to instruct further refactoring operations.
While the authors have carefully described their design decisions and discuss potential success factors for acceptance among developers, an evaluation of the bot is yet missing.

\cite{Wessel2018TheProjects} analyzed 351 Open-Source projects and found that 93 (26\%) use bots which complement other developers' work. The authors interviewed project maintainers to investigate, among other things, how contributors and integrators perceive bots during the pull request submission process. Most respondents perceived bots as helpful for most of the tasks and more than 90\% of them highlighted the relevance of quality assurance tasks. However, among 14 identified bots for code or pull request review there was no bot that automatically provides refactoring suggestions.

\cite{spence2014welcoming} conducted an experiment in which participants were either told that they are going to communicate with a human or that their interaction partner is going to be a robot.
They then measured the uncertainty about the upcoming interaction, anticipated interpersonal liking, and social presence.
\cite{spence2014welcoming} found that participants who believed they were to communicate with another person had higher expectations of liking, lower levels of uncertainty, and higher expectations of social presence than those who believed they were to communicate with a robot.

Related to the liking of bots compared to humans,~\cite{Murgia2016AmongWebsites} have found in an experiment with developers on Stack Overflow that an answer bot is perceived significantly more negatively when it reveals its identity as a bot.
Developers rated the answers of the supposed human as more positive compared to the identical bot, which revealed its identity.
This attitude towards bots could also have a decisive influence in our evaluation study.
We address this in particular with RQ2.

\cite{marcilio2019automatically} evaluated the acceptance of automatic refactoring proposals.
They fixed code smells in 12 projects and proposed 920 fixes in 38 pull requests. 84\% of the pull requests were accepted, 95\% of them without modifications.
The code changes were performed automatically, but proposed manually and only in a short period of time to projects that had previously expressed interest. The reviews of the maintainers and correction requests were also responded to manually.
This differs from the continuous work with an autonomous bot, which submits this kind of code changes and is limited in its interaction possibilities.

\section{\uppercase{Methods}}
\label{sec:methods}

To answer the research questions we had the Refactoring-Bot work in a student software development project over a period of 41 days and without manual intervention of the  authors.
After 11 days, intermediate interviews were conducted with the developers who had interacted with the bot during that time, and based on the interview results, we modified the parameters of the Refactoring-Bot for the remaining 30 days.
After 41 days the final interviews with all project participants took place. 

\subsection{Participants and Project}

The team consisted of 11 bachelor students of Software Engineering, of whom eight were male and three were female.
As part of their studies, they must participate in a six-month joint software project.
In this case, they had to further develop an existing code base with about 18k SLOC.
During the first two months the students were introduced to the project and introduced each other to technologies and methods.
This was followed by the development period, during which the bot was also introduced. 
Students were in the fourth or fifth semester of their studies and had already developed software together in a smaller team in the past.

The backend of the software to be developed was written in Java and the code was hosted as a private GitLab project.
We use the term \enquote{pull request} throughout the paper, meaning suggestions for code changes, although GitLab's terminology uses the term \enquote{merge request}.
The functionality is the same.

\subsection{Research Design}
\label{subsec:researchdesign}

At the beginning of the six-month project, we asked the students if they would agree to the introduction of a refactoring bot to the team at a later date as part of a study.
About two months later, we informed the team that the bot would from now on help to improve the code quality and that we would ask them for their feedback later.

The first phase began, in which we had the bot create exactly one pull request per day at 3pm.
The bot also checked every minute to see if there were any new comments on its pull requests that it had to respond to.
After 11 days, we conducted semi-structured interviews with the only two participants who had interacted with the bot up to that point to capture the first impression of the participants and collect optimization suggestions for the second and longer phase of the study.

The findings from the 45-minute interim interview with both participants gave us initial answers to the research questions and a few of the suggestions for improvement made by the interviewees could be implemented immediately before the start of the second phase.
From then on the bot created pull requests every Tuesday and Thursday between 9am and 6pm every half hour, if the number of already open pull requests by the bot was less than four.
The timing can be explained by the fact that the team meetings always took place on Tuesdays and Thursdays and that it was assumed that all team members would then be on-site and could better focus on the pull requests during this period.

\begin{figure*}
  \centering
  \includegraphics[clip, trim=1.5cm 22.4cm 1.8cm 1.1cm, width=0.99\textwidth, cfbox=lightgray!50!white 0.5pt 0.5pt]{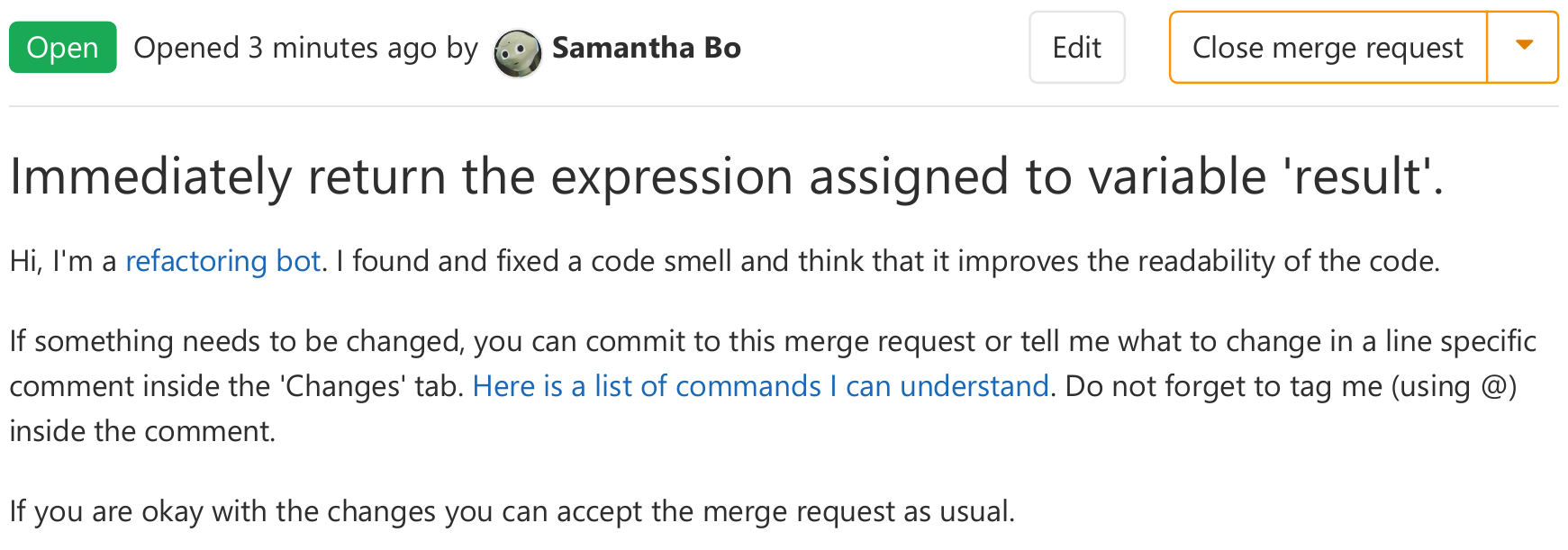}
  \caption{Description of a pull request proposed by the Refactoring-Bot, \textit{Samantha Bo}.}
  \label{fig:examplePR}
\end{figure*}

After the first phase, we also slightly changed the description of the pull requests to include a link to the list of available commands and interaction options.
This was requested by the participants in the interviews.
An example pull request description created by the bot is shown in Figure~\ref{fig:examplePR}.
The team was informed about all changes before the second phase started.

After 41 days the operation of the bot ended and we conducted semi-structured interviews again.
This time nine people took part, even though not all of them were proven to have reviewed a pull request.
However, to evaluate the acceptance of the bot, it is equally important to discuss the attitude and experience of those who consciously or unconsciously did not interact with the bot.
On average, the interviews lasted 20 minutes.

\subsection{Interview Guideline}

We conducted the interim interview and the final interviews following the same procedure.
Participants were invited and informed about the background of the interview and then had the choice to sign a consent form or not to participate.
The interviews were recorded, then transcribed, manually analysed, translated into English and summarised with regard to answering the research questions.

During the interview process, we were guided by an interview guideline.
To not interrupt the flow of the conversation, we adapted the order of the questions to the respective answers.
Occasionally the interviewees also came back to previously asked questions at a later time and added something to their comments.

The contents of the interview guidelines only differed slightly between the interim interview and the final interviews and both served to answer the three research questions.
While the interim interview was about gaining a first impression and starting into the second phase with small and immediately implementable optimizations, these changes were specifically addressed in the final interviews.
The interview guidelines were structured as follows:\\
The first question encouraged the interviewees to reflect openly on their interaction with the bot.
A second question followed and asked directly how the developers felt about the bot's behaviour and its pull requests in terms of quality, usefulness, frequency and timing.
In the final interviews, we mentioned as a third point that we had experimented during the study on frequency and timing and we asked how the study participants perceived this and which variant they would prefer.
We also pointed out the different observations in the two phases (see section~\ref{subsec:acceptance}) and asked the participants for their opinion.
With the answers of the participants to these three questions, we have gained the most insights to answer RQ1.
The fourth question served to answer RQ2 and aimed at how the participants perceived the pull requests of the bot compared to those of human developers and whether they experienced a difference in trust, for example.
The fifth and last question was derived from RQ3, whether the interviewee wanted to use the bot in the future and if so, how the ideal refactoring bot would look like in their future team.

\section{\uppercase{Results}}
\label{sec:results}

In the following we describe the results of our study, grouped by each research question.

\subsection{Perception of the Bot (RQ1)}

\begin{quote}
    \enquote{The problem with our IDE is that it points to bad things, but some developers don't want to fix those things or forget about it or are simply too lazy. And the bot would always do it. That's a real benefit.} -- P02
\end{quote}

\noindent The team members perceived the bot and its refactoring suggestions as useful and repeatedly stated that it offered additional value because it tirelessly improved code quality, which a developer may not be able to do because other things seem to be more important.
The bot would break down a big block of refactoring work and suggest refactoring from time to time while at the same time working on new functionality would be possible.
The bot would also behave unobtrusively.\\
One participant was amused that the first pull request of the bot changed code that the team had never touched.
Otherwise, the quality of the suggestions was found to be good and correct.

Overall, the bot was not often communicated with via comments in pull requests.
However, one participant noticed the limited interaction possibilities negatively, because he expected the Refactoring-Bot to have simple communication skills he knew from chat bots.
In the way the bot describes itself, one could assume that it is possible to talk to it in the same way as with an intelligent virtual assistant.
Another participant wished that one could ask the bot why exactly it made a particular change as part of its refactoring.
For example, it could communicate that it had to change method calls as a result of removing an unused parameter from a signature.
One could also better understand in larger refactorings why something was implemented in a certain way and perhaps even learn something.

Although we did not ask for it, two interviewees commented that they liked the bots name and profile picture.

\vspace{0.3cm}
\noindent
\fbox{%
    \parbox{\columnwidth}{%
        \textbf{RQ1.} The bot was perceived as a useful and unobtrusive contributor. Interaction possibilities should be extended to better understand specific changes and simple chatbot functionalities would be expected.
    }%
}
\vspace{0cm}

\subsection{Acceptance of the Bot (RQ2)}
\label{subsec:acceptance}

\begin{quote}
    \enquote{People also make mistakes sometimes, and we have proven this often enough.} -- P04
\end{quote}

\begin{figure*}
  \centering
  \includegraphics[clip, trim=0cm 1.8cm 0cm 0cm, width=1\textwidth]{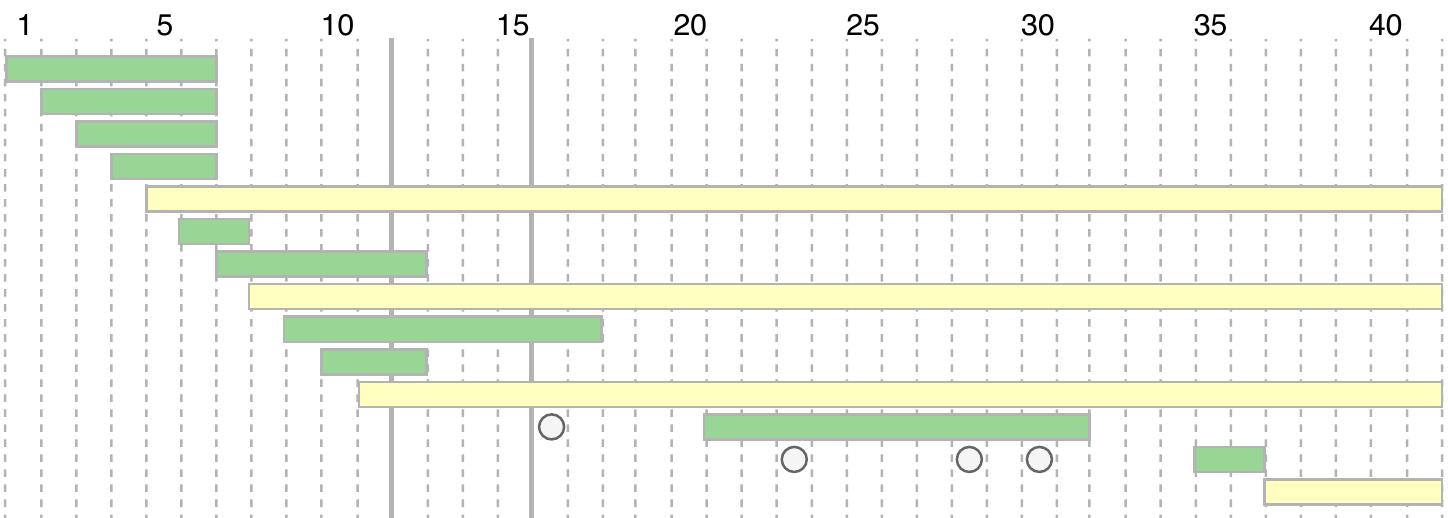}
  \caption{Pull requests created by the bot during the 41-day study period. Two solid vertical lines mark the end of the first phase and the beginning of the second phase. Each green bar represents the time frame from the creation of a pull request to its acceptance. No pull requests were rejected. Yellow bars represent pull requests that were neither accepted nor rejected by the end of the study period. Grey circles stand for times when the bot was prevented from creating a new pull request by the limit of four simultaneously open ones.}
  \label{fig:overviewPRs}
\end{figure*}

\noindent Before we answer the question of whether developers are more or less critical about the bot's contributions than those of humans, we investigate the overall acceptance rate of its pull requests.
Figure~\ref{fig:overviewPRs} shows that a total of 14 pull requests were created, of which eleven were created in the first phase and three in the second phase, which started on day 16.
Ten pull requests were accepted (71.4\%) and four were still open at the end of the study period (28.6\%).
In two of them, the bot removed an unused method parameter that actually should have been used in the method body.
The bot unknowingly pointed out a defect to the developers that could not be corrected in the short term.
They therefore intentionally left the pull requests open as a reminder.

As described in~\ref{subsec:researchdesign}, during the first eleven days one pull request was created each day.
After that, pull requests were created every half hour on Tuesdays and Thursdays, as long as less than four were open at the same time.
While on average two accepted pull requests per week and no rejections indicate a general acceptance of the Refactoring-Bot, the change after the first phase in the frequency of suggestions and the limitation of simultaneously open ones led to a noticeable decrease in the number of processed pull requests.
As a reminder, these changes were implemented based on the responses of the team members from the interim interviews.
When we addressed these changes in the final interviews, we could observe different reactions.
Two of the participants did not notice the changes at all.
Others liked the idea of proposing pull requests when developers are on site, and the idea of having a limit, since developers could more easily motivate themselves to work through only a few open pull requests and it would be easier to keep up with the reviews.
However, the majority of respondents did not consider this change useful in retrospect.
Even the participants of the intermediate interview, who explicitly expressed their preference for limiting the bot, changed their views to optimize the effectiveness of the bot.
If the bot had much to improve, it should also make many suggestions for improvement.
As soon as many pull requests accumulate in the list, the team would be interested in someone taking care of these so that their own pull requests can be found.

The change in parameters revealed the underlying problem that only few team members felt responsible for reviewing the bot's pull requests.
In the interviews, we learned that it was also difficult for human developers to find reviewers for their own changes.
Completing one's own tasks was a higher priority for team members for various reasons.
The bot simply had the disadvantage of not being able to directly contact the other team members and persuade them to review its changes, just as the other team members did for their own pull requests.
In addition, some team members were inexperienced with the version control system and did not dare to merge into the master branch because they were afraid to break something.

In the end, however, most of the bot's pull requests have been accepted.
We asked specifically whether it made a difference for the developers to review the changes of a bot compared to those of a human, for example in terms of trust.
The interviewees replied that, in general, it makes no difference whether the pull request comes from a bot or a human and that this attitude has not changed over time.
The bot would suggest simple changes that the team could be sure would not affect functionality.
And both, human and bot, could sometimes make mistakes.

Participants all felt that it was more important to look at the content of a pull request and to review the content thoroughly.
However, some interviewees also noted that their way of approaching the review would depend on their previous experience with the person or bot who proposed the changes.
\enquote{There are people where I look at the functionality and do not have to worry about code quality. But then there are some people for which I have to have a closer look at their pull requests, because of my recent experience with their work} -- P07.
The same would apply to the Refactoring-Bot.
Furthermore, the changes of a bot would be reviewed with a different focus.
They would know the contributor was a bot that, in turn, would not know what the original code was for.
Therefore, they primarily checked the code changes for logical errors.
The only danger they saw was that the bot might suggest simple refactorings for too long and at some point will propose something riskier that would then not be reviewed sufficiently well.

\vspace{0.2cm}
\noindent
\fbox{%
    \parbox{\columnwidth-0.25cm}{%
        \textbf{RQ2.} Ten out of 14 pull requests were accepted, four were still open at the end of the study period. Only few people felt responsible for reviewing pull requests. The bot was disadvantaged because it could not approach individual team members directly and ask them for a review. However, the participants were neither more nor less critical about the bot than about their human teammates. The previous experience with the contributor was decisive for the thoroughness of the review and changes of the bot were mostly examined for logical errors.
    }%
}
\vspace{0cm}

\subsection{The Ideal Refactoring Bot (RQ3)}

\begin{quote}
    \enquote{Later, in professional life, I am sure someone will take care of it.} -- P08
\end{quote}

\noindent All interviewees explicitly responded that they would like to use the bot in a future project.
The answers to the question what the ideal refactoring bot would look like can be summarized in three points.

First, the frequency of pull requests should be carefully evaluated.
Some consider a fixed limit of open pull requests to be useful, but would probably have set it higher than four and in relation to the number of team members.
Others generally argued against a limit, since these were small changes and the code quality could only be improved effectively without a limit.

Second, there was the request that the changes of the bot should be grouped in a meaningful way, so that less pull requests but bigger ones are created.
At the beginning smaller pull requests were good to get used to the bot.
Later, however, grouping changes has the advantage that fewer refactoring commits would appear in the history and it would be more worthwhile to test the code of the pull request before merging it into the master branch.

By far the most frequently raised point was that one or more people within the team should be made responsible for reviewing the bot's pull requests in a coordinated way.
Respondents could also imagine the bot itself assigning someone to review, at best the developer who is familiar with the part of the code or even the one who caused the code smell.
In any case, it would be important to find a responsible person, and some of the team members also think that this would be easier in a more professional environment.

\vspace{0.3cm}
\noindent
\fbox{%
    \parbox{\columnwidth}{%
        \textbf{RQ3.} The participants of the study were all in favour of using the bot in a future project. It was controversial how often the bot should make suggestions. Participants considered it appropriate to group changes into fewer and slightly larger pull requests, which again would affect the preferred frequency of the bot's suggestions. They agreed that the successful operation of the bot also required someone to feel responsible for reviewing the pull requests.
    }%
}

\section{\uppercase{Discussion}}
\label{sec:discussion}

In the following, we compare our results with those from existing literature, describe the limitations of our study and complete the chapter with implications that raise interesting questions for future work.

\subsection{Results}

We saw that the bot was accepted by the developers, and they explicitly stated that they were neither more nor less critical in reviewing the contributions of the bot compared to those of humans.
This is contrary to the results of~\cite{Murgia2016AmongWebsites}, who had observed a clear negative bias against their answer bot on Stack Overflow.

The approval of 71.4\% of the proposed pull requests and the rejection of none at the same time confirms the general acceptance and is in line with the results of~\cite{marcilio2019automatically} that such code changes are accepted by developers.

Social aspects of the contributor played a role in recent studies~\citep[e.g.]{terrell2017gender,Ford:2019:BeyondCode} and we can confirm that there are at least a few consciously perceived differences from the developer's point of view as to whether the contributor is human or bot.
In the latter case, the proposed changes were mostly reviewed for logic errors.
Furthermore, two participants commented unsolicited that they liked the profile picture and the name of the bot, which confirms the findings to the extent that social aspects are actually perceived.
We also found that previous experience with the bot plays a role in evaluating its code changes.
According to the participants, however, this is no different with humans.

Finally, we have seen that there is potential for improvement in the implementation of the bot.
\cite{Wessel2018TheProjects} analyzed the usage of bots in open source projects.
Most often respondents in their study proposed to make bots smarter and enhance user interaction.
Additionally, they mentioned the lack of information on how to interact with them.
This is consistent with our findings that developers were unsure how to communicate with the bot and expected it to have at least the capabilities of a simple chat bot.
Additionally, the most common suggestion to improve bots in their study from an integrator perspective was that \textit{notification and awareness} should be improved.
This was necessary, for example, to remind developers of unresolved issues.

Improving awareness, especially the sense of responsibility for the bot's pull requests, might be one of the most important insights for the successful use of the Refactoring-Bot.
As long as manual intervention by developers is necessary and the intrinsic motivation of individuals is not high enough, this poses a threat to the acceptance of such maintenance bots.
One explanation could be the \textit{diffusion of responsibility}, a sociopsychological phenomenon in which a person may feel less responsible for actions or inactivity when others are present~\citep{kassin2019social}.

In our scenario, all participants found the bot useful and understood its value.
Because everyone could have reviewed the pull requests, few people might have felt responsible for it.
Of the nine team members interviewed, six stated that they relied on a few specific people because they had also reviewed most of the pull requests of the other team members.

\subsection{Limitations}

The results of this study should be seen in the light of some limitations.
First, the participants in our study were relatively inexperienced in developing software when compared to professional software developers.
This was particularly apparent from the participants' responses that some were uncertain about processing pull requests.
In addition, the low sense of responsibility for reviewing pull requests could be a limitation that is generally associated with the study of students.
This might be different in a team of professionals.

Then, the development time of a software project is typically longer than the time the bot was deployed in our study.
We were limited by the length of the development phase in the student project and would have preferred to deploy the bot for a longer period of time.
The long-term use of the bot needs to be investigated in future studies, as we do not know whether perception and acceptance will improve or deteriorate over time and which design-critical aspects result from the long-term use.\\
In addition, the limited duration of the study did not allow us to wait for the open pull requests from the first phase to be processed.
As a result, four pull requests were already open at the beginning of the second phase, thus limiting the comparability of the two phases.
Since we mainly made general statements about the perception and acceptance of the bot over the entire period, we do not consider the core findings to be affected by this.
On the contrary, we only came to an essential understanding of responsibility issues because these pull requests from the first phase limited the further work of the bot.

The choice of the bot used in the study and its current state of implementation also have limitations.
The Refactoring-Bot currently only supports comparatively simple refactorings.
The review behavior as well as the requirements for frequency and size of pull requests can change with increasing complexity of the refactorings.
However, a look at the code smells reported by static code analysis tools such as SonarQube shows that most of them require rather uncomplicated code improvements.\\
In general, the choice of a refactoring bot and the context in which it was operated could be a reason for some of the observed differences with related work, for example with that of~\cite{Murgia2016AmongWebsites}, in which an answer bot was evaluated on Stack Overflow, where the participating developers could assume interaction only with other human developers.
This deviation from expectation could then have led to the negative bias observed in the study~\citep{spence2014welcoming}.
In contrast, the participation of bots in software projects is becoming increasingly common~\citep{Wessel2018TheProjects} and may therefore generate greater acceptance, as was the case in our study.

\subsection{Implications}
\label{subsec:implications}

\subsubsection*{Developers may accept bots -- even if they are identified as such}

We saw that the bot in our study was accepted by the developers and even to the extent that the review of the bot's contributions was no more critical than that of human developers.
This should create confidence in the bot community that the automation of software engineering tasks via a bot interface generally has the potential to be accepted by developers.
Since this has been different in previous studies in other contexts, the impact of a bot's operating context on its design and acceptance should be investigated more closely.

\subsubsection*{Bots need to be process-sensitive}

In addition to context-dependent design, the question arose as to how process-sensitive a bot must be.
In this case, the bot did not correspond to the process of the team, which includes that the entity contributing changes must find a reviewer itself.
This could easily be resolved by automatically assigning a suitable reviewer to the bot's pull requests.
However, we do not know if this would actually be enough, or if the underlying problem is that few people feel responsible for dealing with the bot.
The smarter advice might be to make it difficult to ignore the bot when designing a demonstrably useful bot whose effectiveness depends on human intervention.

\subsubsection*{Bots need to be smart}

Finally, bots must be intelligent enough to provide developers with a satisfying experience of interaction.
This includes describing exactly how to interact with them and that there exists a basic repertoire of chat bot functionalities that are common in many applications nowadays.
The bot should also be able to explain its activities upon request.
In our case, the requirement arose to be able to explain individual code changes of a refactoring.

\section{\uppercase{Conclusion}}
\label{sec:conclusion}

A refactoring bot was developed in the past to overcome the drawbacks of removing code smells manually and to make the process more efficient and effective.
As similar bots did not always meet with high acceptance in the context of software engineering and since we have little qualitative knowledge about the perception and acceptance of development bots in general, we conducted a qualitative study deploying a refactoring bot to investigate how it is perceived and accepted in a software development team.

In semi-structured interviews we found that the bot is perceived as a useful and unobtrusive contributor.
Its contributions are not reviewed more critically than those of human developers but are more intensively analyzed for logical errors.
Although all team members intend to continue using the bot, only a few felt responsible for it during the study.

The results have implications for the design of bots to ensure their successful use.
Even a useful bot will be ignored if its perceived benefit for the individual is not great enough and there is a diffusion of responsibility in the team.
Developers accept bots, but they must be smart and adapted to the context and process of the development team.

Future work could explore which context variables predict the acceptance of a bot and investigate more bots qualitatively.
Their long-term use should be investigated along with how the developers' perception changes over time.
In the context of refactoring bots, the effects of proposing more complex and diverse refactorings on the behavior of developers should be studied.
The findings would also be of interest to other maintenance bots which regularly propose very similar changes, possibly leading to a kind of review blindness.
Finally, the results can be understood as an indication that people appreciated that a socially unpleasant task was taken over by the bot, which was to remind people that code quality needs to be improved.
It would be interesting to further explore the potential of this type of bot contribution.

\bibliographystyle{apalike}
{\small
\bibliography{main}}

\end{document}